# *µ*TRISTAN and LHC/Tevatron/FCC/SppC Based Antimuon-Hadron Colliders

Dilara Akturk[1,*], Burak Dagli[1], Bora Ketenoglu[2], Arif Ozturk[1,3], Saleh Sultansoy[1,4]

[1]TOBB University of Economics and Technology, Ankara, Türkiye
[2]Department of Engineering Physics, Ankara University, Ankara, Türkiye
[3]Turkish Accelerator Radiation Laboratory (TARLA), Ankara, Türkiye
[4]ANAS Institute of Physics, Baku, Azerbaijan
[*]email: d.akturk@etu.edu.tr

**Abstract**

Recently, the construction of $\mu^+e^-$ and $\mu^+\mu^+$ colliders, *µ*TRISTAN, at KEK has been proposed. We argue that the construction of a similar $\mu^+$ ring tangential to LHC/Tevatron/FCC/SppC will give an opportunity to realize $\mu^+p$ and $\mu^+A$ collisions at multi-TeV scale center-of-mass energies. In this paper the main parameters of proposed colliders have been studied. It is shown that sufficiently high luminosities can be achieved for all proposals under consideration: *L* exceeds $10^{33}$ cm$^{-2}$s$^{-1}$ for $\mu^+p$ colliders and $10^{30}$ cm$^{-2}$s$^{-1}$ for $\mu^+A$ colliders. Certainly, proposed colliders will provide huge potential for both SM (especially QCD basics) and BSM physics searches.

## 1. Introduction

The electro-weak part of the Standard Model (SM) was completed with the discovery of Higgs boson [1, 2]. Mass and mixing patterns of the SM fermions is another story to be solved by Beyond the Standard Model (BSM) physics. Considering QCD part of the SM, there are still big gaps: confinement hypothesis has still not been proved at QCD basics, hadronization and nuclearization phenomena are not clearly understood. For these reasons, energy-frontier lepton-hadron colliders are required, since lepton colliders can not clarify this phenomena, while hadron colliders have huge background.

It is worth pointing out that lepton-hadron collisions played a crucial role in our understanding of the deep structure of matter:

- proton form-factors were first observed in electron scattering experiments [3, 4]

- quarks were first observed at SLAC deep inelastic electron scattering experiments [5, 6]

- EMC effect was observed at CERN in deep inelastic muon scattering experiments [7] and so on.

HERA [8], the first and still unique electron-proton collider, further explored structure of protons and provided parton distribution functions (PDFs) for Tevatron and LHC. TeV (or multi-TeV) scale lepton-hadron colliders are crucial for clarifying basics of QCD, which is responsible for 98% of mass of the visible part of our Universe. Unfortunately, HERA could not cover the region of sufficiently small *x*-Bjorken ($< 10^{-4}$) at high $Q^2$ ($> 10$ GeV$^2$), which is very important for understanding QCD basics. In order to investigate this region, lepton-hadron colliders with higher center-of-mass energies are required. Today, linac-ring type *eh* colliders seem to be the most realistic way to TeV scale in *lh* collisions (see review [9]), and LH*e*C [10] is the most advanced proposal. Concerning *e*RHIC [11], its *ep* center-of-mass energy is three times lower than HERA (main advantages of the *e*RHIC are *eA* collisions and polarized beams).

Besides, contruction of TeV energy lepton-hadron colliders is mandatory to provide PDFs for adequate interpretation of forthcoming data from HL/HE-LHC [12, 13] and FCC/SppC [14, 15] as HERA provided that for Tevatron and LHC in the last two decades. Finally, energy-frontier lepton-hadron colliders are advantageous for investigation of a number of BSM phenomena [16].

Today, there are five known ways to reach multi-TeV scale: proton colliders (LHC, FCC, SppC), linear electron-positron colliders (CLIC [17], PWFA-LC [18]), muon colliders [19], linac-ring type electron-proton colliders (see review [9] and references therein, as well as, [20] and [21] for FCC and SppC based lepton-hadron colliders, respectively) and ring type muon-proton colliders (see review [22] and references therein). Below, we present a correlation between colliding beams and colliding schemes in energy-frontier aspect.

Energy frontier colliders: colliding beams vs. collider types.

| Colliders | Ring-Ring | Linac-Linac | Linac-Ring |
|---|---|---|---|
| Hadron | + | | |
| Lepton ($e^-e^+$) | | + | |
| Lepton ($\mu^-\mu^+$) | + | | |
| Lepton-hadron (*eh*) | | | + |
| Lepton-hadron ($\mu h$) | + | | |
| Photon-hadron | | | + |

Even though muon-proton colliders appear to be in fifth place at first glance, they may attain second place after proton colliders. The reason for this is that the cooling problem of the $\mu^+$ beam has been solved at JPARC [23]. When it comes to the realization of muon colliders, the muons also need to be cooled, however even the proof of principle for the ionisation cooling of $\mu^-$ beam is planned for the 2030s [24].

Recently, the $\mu$TRISTAN [25] project that includes $\mu^+e^-$ ($\sqrt{s}$ = 346 GeV, $L = 4.6 \times 10^{33}$ cm$^{-2}$s$^{-1}$) and $\mu^+\mu^+$ ($\sqrt{s}$ = 2 TeV, $L = 5.7 \times 10^{32}$ cm$^{-2}$s$^{-1}$) colllider options have been put forward based on $\mu^+$ cooling. Subsequently, we proposed the construction of FCC and SppC-based $\mu^+e^-$ and $\mu^+p$ colliders [26] using the $\mu^+$ beam parameters of the $\mu$TRISTAN project.

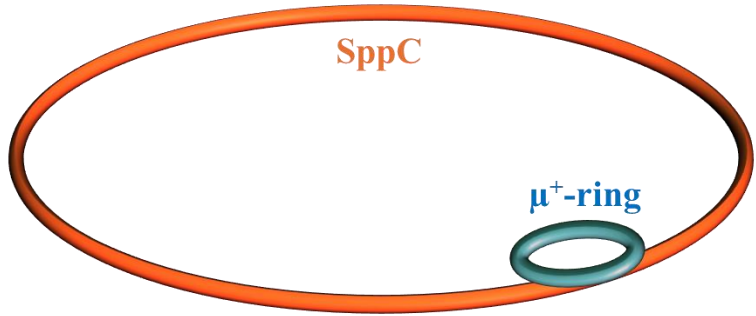

**Figure 1.** Schematic view of antimuon-hadron collider

In this paper, we consider the main parameters of μTRISTAN and LHC/Tevatron/FCC/SppC based $\mu^+p$ and $\mu^+A$ colliders (schematic view of proposed $\mu^+p/A$ collider is illustrated in Figure 1). These parameters are evaluated for each proposed collider in Section 2. Comments on physics search potential of $\mu^+p$ colliders are given in Section 3. Finally, in Section 4 we present our conclusion and recommendations.

## 2. Main parameters of $\mu^+p$ and $\mu^+A$ Colliders

Main parameters of colliders under consideration are evaluated through the software AloHEP (A luminosity optimizer for High Energy Physics) [27-29], which calculates center-of-mass energy ($\sqrt{s}$), luminosity (L), transverse beam sizes ($\sigma_x$, $\sigma_y$) and beam-beam tuneshift ($\xi_x$, $\xi_y$) parameters. This software simulates the collision of bunches which consist of normal-distributed macroparticles, considering hourglass effect, crossing angle and beam-beam interactions in the collision region. It was developed several years ago for estimation of main parameters of linac-ring type *ep* colliders. Later, AloHEP was upgraded for all types of colliders (linear, circular, and linac-ring) and colliding beams (electron, positron, muon, proton, and nuclei).

### 2.1. μTRISTAN based $\mu^+p$ and $\mu^+A$ Colliders

Antimuon beam parameters of the μTRISTAN $\mu^+e^-$ proposal [25] are presented in Table 1. These parameters are used for all collider options, which are considered in this paper. As mentioned in introduction μTRISTAN proposal also includes $\mu^+\mu^+$ collider option. For the $\mu^+\mu^+$ collider, the $\mu^+$ beam is split into two so that each beam has 20 bunches, while the $\mu^+e^-$ collider can use all 40 bunches.

**Table 1.** Main parameters of μTRISTAN's $\mu^+$ beam

| Parameter | Antimuon |
|---|---|
| Number of Particle per Bunch [$10^{10}$] | 1.4 |
| Beam Energy [GeV] | 1000 |
| Horizontal β Function @ IP [cm] | 3 |
| Vertical β Function @ IP [cm] | 0.7 |
| Bunch Length [mm] | 2 |
| Norm. Horizontal Emittance [μm] | 4 |
| Norm. Vertical Emittance [μm] | 4 |
| Number of Bunches per Ring | 40 |
| Collision Frequency [MHz] | 4 |
| Circumference [km] | 3 |

μTRISTAN based $\mu^+p$ collider was proposed in our previous paper [26], where the parameters of ERL60 upgraded FCC proton beam [14] modified for 3 km tunnel have been used. Here, we used the modified parameters of proton and lead beam from HL-LHC ERL60 upgraded version (see Tables 2.11 and 2.13 in Ref. [14]). These modified parameters are presented in Table 2.

Table 2. ERL60 upgraded HL-LHC proton and lead parameters for 3 km tunnel

| Parameter | Proton | | Lead (Pb) | |
|---|---|---|---|---|
| Bending Magnetic Field | 8 Tesla | 16 Tesla | 8 Tesla | 16 Tesla |
| Number of Particle per Bunch [$10^{10}$] | 22 | | 0.018 | |
| Beam Energy [TeV] | 0.85 | 1.7 | 177 | 354 |
| β Function @ IP [cm] | 7 | | 7 | |
| Norm. Emittance [μm] | 2 | | 1.5 | |
| Number of Bunches per Ring | 40 | | | |
| Revolution Frequency [kHz] | 100 | | | |
| Circumference [km] | 3 | | | |

By entering the values from Table 1 and Table 2 into AloHEP, the parameters of the $\mu^+p$ and $\mu^+Pb$ colliders shown in Tables 3 and 4 are obtained.

Table 3. Main parameters of the $\mu$TRISTAN based $\mu^+p$ colliders

| Parameter | 8 Tesla | | 16 Tesla | |
|---|---|---|---|---|
| $\sqrt{s}$ [TeV] | 1.84 | | 2.61 | |
| L [$10^{32}$ cm$^{-2}$s$^{-1}$] | 6.3 | | 13 | |
| Parameter | Proton | Muon | Proton | Muon |
| σ$_{x,y}$ [μm] | 12 | | 8.8 | |
| Tuneshift ($\xi_x$) [$10^{-4}$] | 8.6 | 600 | 8.6 | 600 |

It is seen that muTRISTAN based $\mu^+p$ colliders have higher center-of-mass energies than LHeC [10], which is important for clarification of QCD basics.

Table 4. Main parameters of the $\mu$TRISTAN based $\mu^+Pb$ colliders

| Parameter | 8 Tesla | | 16 Tesla | |
|---|---|---|---|---|
| $\sqrt{s}$ [TeV] | 26.6 | | 37.6 | |
| L [$10^{30}$ cm$^{-2}$s$^{-1}$] | 0.70 | | 1.4 | |
| Parameter | Pb | Muon | Pb | Muon |
| σ$_{x,y}$ [μm] | 11 | | 7.6 | |
| Tuneshift ($\xi_{x,y}$) [$10^{-2}$] | 3.7 | 0.40 | 3.7 | 0.40 |

## 2.2. HL-LHC based $\mu^+p$ and $\mu^+A$ Colliders

In this subsection, for hadron beams, we used HL-LHC $p/Pb$ beam parameters upgraded for ERL60-based $ep$ collider (Tables 2.11 and 2.13 in Ref. [14]). These parameters are presented in Table 5.

Table 5. ERL60 upgraded proton and lead parameters

| Parameter | Proton | Pb |
|---|---|---|
| Number of Particle per Bunch [$10^{10}$] | 22 | 0.018 |
| Beam Energy [TeV] | 7 | 574 |
| β Function @ IP [cm] | 7 | 7 |
| Norm. Emittance [μm] | 2 | 1.5 |
| Number of Bunches per Ring | 2760 | 1200 |
| Revolution Frequency [Hz] | 11245 | |
| Circumference [km] | 26.7 | |

By implementing the parameters of antimuon beam from Table 1 and *p/Pb* beams from Table 5 into AloHEP, we obtained the results for main parameters of HL-LHC $\mu^+ p$ and $\mu^+ Pb$ colliders which are given in Tables 6 and 7, respectively.

Table 6. Main parameters of the HL-LHC based $\mu^+ p$ collider

| Parameter | HL-LHC | |
|---|---|---|
| $\sqrt{s}$ [TeV] | 5.29 | |
| L [$10^{33}$ cm$^{-2}$s$^{-1}$] | 5.2 | |
| Parameter | Proton | Muon |
| $\sigma_{x,y}$ [μm] | 4.3 | 4.3 |
| Tuneshift ($\xi_{x,y}$) [$10^{-4}$] | 8.6 | 600 |

Table 7. Main parameters of the HL-LHC based $\mu^+ Pb$ colliders

| Parameter | HL-LHC | |
|---|---|---|
| $\sqrt{s}$ [TeV] | 47.9 | |
| L [$10^{30}$ cm$^{-2}$s$^{-1}$] | 2.3 | |
| Parameter | Pb | Muon |
| $\sigma_{x,y}$ [μm] | 6.0 | 6.0 |
| Tuneshift ($\xi_{x,y}$) [$10^{-4}$] | 370 | 40 |

One can see that center-of-mass energies are 4 times higher than LH*e*C, while luminosities are comparable. Therefore, HL-LHC based $\mu^+ p$ collider will give opportunity to investigate an order smaller values of *x* Björken.

### 2.3. Tevatron based $\mu^+ p$ and $\mu^+ A$ Colliders

The parameters of proton and Pb beams obtained by upgrade of Table 5 for Tevatron energies are given in Table 8.

**Table 8.** Tevatron proton and lead parameters

| Parameter | Proton | | Lead (*Pb*) | |
|---|---|---|---|---|
| Bending Magnetic Field | 8 Tesla | 16 Tesla | 8 Tesla | 16 Tesla |
| Number of Particle per Bunch [$10^{10}$] | 22 | | 0.018 | |
| Beam Energy [TeV] | 1.78 | 3.56 | 370 | 740 |
| β Function @ IP [cm] | 7 | | 7 | |
| Norm. Emittance [μm] | 2 | | 1.5 | |
| Number of Bunches per Ring | 76 | | | |
| Revolution Frequency [Hz] | 47770 | | | |
| Circumference [km] | 6.28 | | | |

By entering the parameters of antimuon beam from Table 1 and *p/Pb* beams from Table 8 into AloHEP, we obtained the results for main parameters of HL-LHC $\mu^+p$ and $\mu^+Pb$ colliders which are given in Tables 9 and 10, respectively.

**Table 9.** Main parameters of the Tevatron based $\mu^+p$ colliders

| Parameter | 8 Tesla | | 16 Tesla | |
|---|---|---|---|---|
| $\sqrt{s}$ [TeV] | 2.66 | | 3.77 | |
| L [$10^{33}$ cm$^{-2}$s$^{-1}$] | 1.2 | | 2.4 | |
| Parameter | Proton | Muon | Proton | Muon |
| $\sigma_{x,y}$ [μm] | 8.6 | | 6.1 | |
| Tuneshift ($\xi_x$) [$10^{-4}$] | 8.6 | 600 | 8.6 | 600 |

**Table 10.** Main parameters of the Tevatron based $\mu^+Pb$ colliders

| Parameter | 8 Tesla | | 16 Tesla | |
|---|---|---|---|---|
| $\sqrt{s}$ [TeV] | 38.5 | | 54.4 | |
| L [$10^{30}$ cm$^{-2}$s$^{-1}$] | 1.3 | | 2.6 | |
| Parameter | Pb | Muon | Pb | Muon |
| $\sigma_{x,y}$ [μm] | 7.4 | 7.4 | 5.3 | 5.3 |
| Tuneshift ($\xi_{x,y}$) [$10^{-2}$] | 3.7 | 0.40 | 3.7 | 0.40 |

It is seen that center-of-mass energies are 2 (for 8 T magnets) and 3 (for 16 T magnets) times higher than LH*e*C, while luminosities are comparable.

**2.4. FCC based $\mu^+p$ and $\mu^+A$ Colliders**

FCC based $\mu^+p$ collider has been considered in [26]. Main parameters of this collider are presented in Table 11.

**Table 11.** Main parameters of the FCC based $\mu^+p$ collider

| Parameter | FCC | |
|---|---|---|
| $\sqrt{s}$ [TeV] | 14.1 | |
| L [$10^{32}$ cm$^{-2}$s$^{-1}$] | 50 | |
| Parameter | Proton | Muon |
| $\sigma_x$ [μm] | 3.6 | |
| $\sigma_y$ [μm] | 2.5 | |
| Tuneshift ($\xi_x$) [$10^{-4}$] | 6.4 | 220 |
| Tuneshift ($\xi_y$) [$10^{-4}$] | 9.2 | 320 |

In Table 12, we present parameters of ERL60 upgraded FCC lead beam (Table 2.13 in Ref. [14]).

**Table 12.** FCC lead beam parameters

| Parameter | Pb |
|---|---|
| Number of Particle per Bunch [$10^{10}$] | 0.018 |
| Beam Energy [TeV] | 4100 |
| β Function @ IP [cm] | 15 |
| Norm. Emittance [μm] | 0.9 |
| Number of Bunches per Ring | 2072 |
| Revolution Frequency [Hz] | 2998 |
| Circumference [km] | 100 |

Implementing parameters from Tables 1 and 12 into AloHEP, we obtain the main parameters of the FCC based $\mu^+Pb$ collider which are presented in Table 13.

**Table 13.** Main parameters of the FCC based $\mu^+Pb$ colliders

| Parameter | FCC | |
|---|---|---|
| $\sqrt{s}$ [TeV] | 128.1 | |
| L [$10^{30}$ cm$^{-2}$s$^{-1}$] | 8.9 | |
| Parameter | Pb | Muon |
| $\sigma_{x,y}$ [μm] | 6.0 | 6.0 |
| Tuneshift ($\xi_{x,y}$) [$10^{-4}$] | 616 | 40 |

One can see that center-of-mass energies are 4 times higher than ERL60 based FCC-*eh*, while luminosities are comparable. Therefore, FCC based $\mu^+p$ collider will give opportunity to investigate an order smaller values of *x* Björken.

## 2.5. SppC based $\mu^+p$ and $\mu^+A$ Colliders

SppC based $\mu^+p$ collider has been considered in [26] where we used proton beam parameters from PDG [30]. In this paper we use proton and lead beam parameters from recent CEPC TDR (see Table A7.2 in [31]).

Table 14. SppC proton and lead beam parameters

| Parameter | proton | Pb |
|---|---|---|
| Number of Particle per Bunch [$10^{10}$] | 15 | 0.18 |
| Beam Energy [TeV] | 37.5 | 3075 |
| β Function @ IP [cm] | 75 | 75 |
| Norm. Emittance [μm] | 2.35 | 0.22 |
| Number of Bunches per Ring | 10080 | 10080 |
| Revolution Frequency [Hz] | 3000 | |
| Circumference [km] | 100 | |

Implementing parameters from Tables 1 and 14 into AloHEP, we obtain the main parameters of the SppC based $\mu^+p$ and $\mu^+Pb$ collider which are presented in Tables 15 and 16, respectively.

Table 15. Main parameters of the SppC based $\mu^+p$ collider

| Parameter | SppC | |
|---|---|---|
| $\sqrt{s}$ [TeV] | 12.3 | |
| L [$10^{33}$ cm$^{-2}$s$^{-1}$] | 1.5 | |
| Parameter | proton | Muon |
| $\sigma_x$ [μm] | 6.6 | 6.6 |
| $\sigma_y$ [μm] | 6.6 | 6.6 |
| Tuneshift ($\xi_x$) [$10^{-4}$] | 7.3 | 409 |
| Tuneshift ($\xi_y$) [$10^{-4}$] | 7.3 | 409 |

Table 16. Main parameters of the SppC based $\mu^+Pb$ collider

| Parameter | SppC | |
|---|---|---|
| $\sqrt{s}$ [TeV] | 111 | |
| L [$10^{31}$ cm$^{-2}$s$^{-1}$] | 7.00 | |
| Parameter | Pb | Muon |
| $\sigma_x$ [μm] | 3.6 | 3.6 |
| $\sigma_y$ [μm] | 3.2 | 3.2 |
| Tuneshift ($\xi_x$) [$10^{-2}$] | 24 | 3.8 |
| Tuneshift ($\xi_y$) [$10^{-2}$] | 26 | 4.2 |

For SppC-based lepton-hadron colliders, the 120 GeV electron beam provided by CEPC is planned to be used (see Appendix 7 in [31]). Using a 1 TeV antimuon beam instead would provide a 3-fold advantage in center-of-mass energy.

## 3. Brief Remarks on Physics Search Potential

Certainly, multi-TeV center-of-mass energy antimuon-hadron colliders have a huge potential for both the SM and BSM physics searches (see for example review articles [32, 33]). Here, we restrict ourselves to the consideration of small $x$-Bjorken region and Higgs boson production.

### 3.1. Small $x_g$

As mentioned in Introduction, investigation of the region of sufficiently small $x$-Bjorken ($<10^{-4}$) at high $Q^2$ ($>10$ GeV$^2$) is very important for understanding QCD basics. For $\mu p$ colliders, the relation between $x$-Bjorken and $Q^2$ is given as $Q^2 = 4x_B E_\mu E_p$. Achievable $x$-Bjorken values at $Q^2 = 100$ GeV$^2$ and $Q^2$ values at $x = 10^{-5}$ for proposed colliders is presented in Table 17.

**Table 17.** Achievable $x$-Bjorken values at $Q^2 = 100$ GeV$^2$ and $Q^2$ values at $x = 10^{-5}$

|  |  | $\sqrt{s}$ [TeV] | $x$-Bjorken | $Q^2$ [GeV$^2$] |
|---|---|---|---|---|
| HL-LHC |  | 5.29 | $3.6 \times 10^{-6}$ | 280 |
| FCC |  | 14.1 | $5.0 \times 10^{-7}$ | 1990 |
| SppC |  | 12.3 | $6.6 \times 10^{-7}$ | 1510 |
| Tevatron | 8 T | 2.66 | $1.4 \times 10^{-5}$ | 71 |
|  | 16 T | 3.77 | $7.0 \times 10^{-6}$ | 142 |
| $\mu$TRISTAN | 8 T | 1.84 | $3.0 \times 10^{-5}$ | 34 |
|  | 16 T | 2.61 | $1.5 \times 10^{-5}$ | 68 |

As can be seen from the table, the proposed colliders will allow detailed exploration of the relevant ($x$-$Q^2$) region.

### 3.2. Higgs Production

Cross-sections for Higgs boson production in $\mu^+ p$ collisions via W and Z fusion processes are presented in Figure 2.

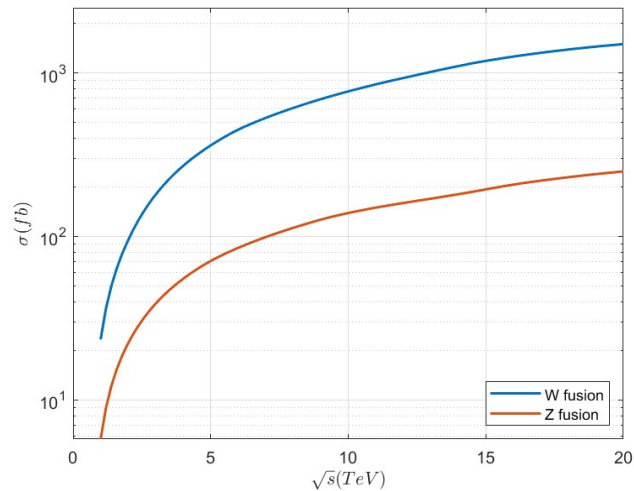

**Figure 2.** Higgs production in $\mu^+ p$ collisions

Number of Higgs bosons produced per working year ($10^7$ s) for colliders under consideration are given in Table 18. For comparison corresponding numbers for LHeC were also included, where the latest LHeC parameters [34] have been used: $E_e$ = 50 GeV, $E_p$ = 7 TeV and L = $1.4\times10^{33}$ cm$^{-2}$s$^{-1}$. It is clearly seen that $\mu^+p$ colliders are much more advantageous than LHeC.

Table 18. Number of Higgs bosons produced per working year

| Colliders | | $\sqrt{S}$ [TeV] | $L_{int}$ [fb$^{-1}$] | $N_{WFus}$ | $N_{ZFus}$ |
|---|---|---|---|---|---|
| LHeC | | 1.18 | 14 | 490 | 120 |
| HL-LHC | | 5.29 | 52 | 20000 | 3900 |
| FCC | | 14.1 | 50 | 55000 | 9000 |
| SppC | | 12.3 | 15 | 14000 | 2400 |
| Tevatron | 8 T | 2.66 | 12 | 1900 | 400 |
| | 16 T | 3.77 | 12 | 3000 | 600 |
| $\mu$TRISTAN | 8 T | 1.84 | 6.3 | 540 | 120 |
| | 16 T | 2.61 | 13 | 2100 | 420 |

It should be mentioned that corresponding proton colliders produce essentially more Higgs bosons. However, $\mu^+p$ colliders can provide an important contribution to determining the properties of Higgs bosons due to the cleaner environment compared to proton colliders. This topic needs to be examined in detail.

## 4. Conclusion

Construction of $\mu$TRISTAN $\mu^+$ ring tangential to existing and proposed hadron colliders will provide the opportunity to realize $\mu^+p$ ($\mu^+A$) colliders with multi-TeV center-of-mass energies at luminosities exceeding $10^{33}$ cm$^{-2}$s$^{-1}$ ($10^{30}$ cm$^{-2}$s$^{-1}$). Obviously, such colliders will essentially enlarge the physics search potential of corresponding hadron colliders for both the SM and BSM phenomena. Therefore, systematic studies of accelerator, detector and physics search aspects of $\mu$TRISTAN/LHC/Tevatron/FCC/SppC based $\mu^+p$ and $\mu^+A$ colliders are necessary for long-term planning of High Energy Physics.

Main parameters of $\mu^+p$ and $\mu^+A$ colliders considered in Section 2 are summarized in Table 19. For comparison, recent parameters of the LHeC $ep$ collider are $\sqrt{S}$ = 1.18 TeV with $L_{ep}$ = $1.4\times10^{33}$ cm$^{-2}$s$^{-1}$ [34], FCC-$ep$ are $\sqrt{S}$ = 3.46 TeV with $L_{ep}$ = $1.5\times10^{34}$ cm$^{-2}$s$^{-1}$ [14] and SppC-$ep$ are $\sqrt{S}$ = 4.2 TeV with $L_{ep}$ = $4.5\times10^{33}$ cm$^{-2}$s$^{-1}$ [15]. It is seen that center-of-mass energies of $\mu^+p$ colliders are essentially higher than that of $ep$ colliders, while the luminosities are of the same order.

**Table 19.** Main parameters of $\mu^+p$ and $\mu^+Pb$ colliders

| Colliders | | | $\sqrt{S}$ [TeV] | L [cm$^{-2}$s$^{-1}$] |
|---|---|---|---|---|
| $\mu p$ | HL-LHC | | 5.29 | 5.2×10$^{33}$ |
| | FCC | | 14.1 | 5.0×10$^{33}$ |
| | SppC | | 12.3 | 1.5×10$^{33}$ |
| | Tevatron | 8 T | 2.66 | 1.2×10$^{33}$ |
| | | 16 T | 3.77 | 1.2×10$^{33}$ |
| | $\mu$TRISTAN | 8 T | 1.84 | 6.3×10$^{32}$ |
| | | 16 T | 2.61 | 13×10$^{32}$ |
| $\mu A$ | HL-LHC | | 47.9 | 2.3×10$^{30}$ |
| | FCC | | 128.1 | 8.9×10$^{30}$ |
| | SppC | | 111 | 7.0×10$^{31}$ |
| | Tevatron | 8 T | 38.5 | 1.3×10$^{30}$ |
| | | 16 T | 54.4 | 2.6×10$^{30}$ |
| | $\mu$TRISTAN | 8 T | 26.6 | 0.70×10$^{30}$ |
| | | 16 T | 37.6 | 1.4×10$^{30}$ |

As mentioned in Section 3, $\mu^+p$ colliders will make essential contributions to the clarification of QCD basics and Higgs boson properties. Concerning BSM physics, these colliders will provide a huge potential for investigation of muon-related new phenomena, namely, excited muon, excited muon neutrino, color octet muon, leptoquarks, contact interactions, etc.

Regarding LHC, FCC and SppC-based lepton-hadron colliders, a two-stage approach can be realized: an electron ring as the first stage followed by a muon ring in the same ring as the second stage. For SppC and FCC, a larger ring could be considered, for example, 15 km long to handle $\mu^+$ beam with 5 TeV energy. In this scenario muon collider with $\sqrt{s}$ = 10 TeV center-of-mass energy can be constructed as last stage if the muon cooling problem is resolved.

Finally, let us emphasize that the antimuon-hadron colliders under consideration can be realized before the muon colliders. The reason is that while $\mu^-$ beams with emittance, which is sufficiently small for the construction of muon colliders, have not yet been achieved, there is an established technology to create a low emittance $\mu^+$ beam by using ultra-cold muons [23]. And let us remember that muon-proton colliders might become the second most effective tool, after proton colliders, for exploring the multi-TeV scale at the constituent level.

**References**

1. G. Aad, B. Abbott, J. Abdallah et al., Phys. Lett. B 716, 1 (2012)

2. S. Chatrchyan, V. Khachatryan, A. M. Sirunyan et al., Phys. Lett. B 716, 30 (2012)

3. R. Hofstadter and R. W. McAllister, Phys. Rev. 98, 217 (1955)

4. R. W. McAllister and R. Hofstadter, Phys. Rev. 102, 851 (1956)

5. M. Breidenbach, J. I. Friedman, H. W. Kendall et al., Phys. Rev. Lett. 23, 935 (1969)

6. E. D. Bloom, D. H. Coward, H. DeStaebler et al., Phys. Rev. Lett. 23, 930 (1969)

7. J.-J. Aubert, G. Bassompierre, K. H. Becks et al., Phys. Lett. B 123, 275 (1983)

8. M. Klein and R. Yoshida, Prog. Part. Nucl. Phys. 61(2), 343-393 (2008)

9. A. N. Akay, H. Karadeniz and S. Sultansoy, Int. J. of Mod. Phys. A 25, 4589 (2010)

10. J. A. Fernandez, C. Adolphsen, A. N. Akay et al., J. Phys. G 39, 075001 (2012)

11. R. Abdul Khalek, A. Accardi, J. Adam et al., Nucl. Phys. A 1026, 122447 (2022)

12. I. Zurbano Fernandez, M. Zobov, A. Zlobin et al., Tech. Rep. CERN-2020-010, CERN, Geneva (2020)

13. A. Abada, M. Abbrescia, S. S. AbdusSalam, et al., Eur. Phys. J. Spec. Top. 228, 1109 (2019)

14. A. Abada, M. Abbrescia, S. S. AbdusSalam et al., Eur. Phys. J. Spec. Top. 228, 755 (2019)

15. CEPC Study Group, arXiv:1809.00285 [physics.acc-ph]

16. S. Sultansoy, A Review of TeV Scale Lepton-Hadron and Photon-Hadron Colliders, in *Proceedings of 2005 Particle Accelerator Conference* (Knoxville, Tennessee, 2005), p. 4329, arXiv: hep-ex/0508020

17. T.K. Charles, P.J. Giansiracusa, T.G. Lucas et al., CERN–2018–010–M, CERN, Geneva (2018)

18. E. Adli, J-P. Delahaye, S.J. Gessner et al., arXiv:1308.1145 [physics.acc-ph]

19. C. Accettura, D. Adams, R. Agarwal et al., Eur. Phys. J. C 83, 864 (2023)

20. Y.C. Acar, A.N. Akay, S. Beser et al., Nucl. Inst. and Meth. A 871, 47 (2017)

21. A. C. Canbay, U. Kaya, B. Ketenoglu et al., Adv. High Energy Phys 2017, 4021493 (2017)

22. B. Ketenoglu, B. Dagli, A. Ozturk and S. Sultansoy, Mod. Phys. Lett. A 37, No. 37n38, 2230013 (2022)

23. Y. Kondo et al., Re-acceleration of ultra cold muon in J-PARC muon facility, in *Proceedings of 9th International Particle Accelerator Conference* (Canada, 2018), p. 6

24. D. Schulte, Project Status and Directions (IMCC and MuCol), *presentation at IMCC and MuCol Annual Meeting*, 12-15 March 2024, CERN

25. Y. Hamada, R. Kitano, R. Matsudo et al., Prog. Theor. Exp. Phys 5, 053B02 (2022)

26. D. Akturk, B. Dagli and S. Sultansoy, Muon ring and FCC-ee / CEPC based antimuon-electron colliders, EPL, to be published, arXiv: 2403.17034 (2024)

27. B. Dagli, S. Sultansoy, B. Ketenoglu et al., Beam-beam simulations for lepton-hadron colliders: Alohep software, in *Proceedings of 12th International Particle Accelerator Conference* (Brazil, 2021), p. 3293

28. https://github.com/yefetu/ALOHEP, retrieved 29th July 2024

29. http://yef.etu.edu.tr/ALOHEP_eng.html, retrieved 29th July 2024

30. V. Shiltsev and F. Zimmermann, Prog. Theor. Exp. Phys 2022, Issue 8, p. 533 (2022)

31. W. Abdallah et al. (CEPC Study Group), arXiv:2312.14363 [physics.acc-ph]

32. K. Cheung and Z. S. Wang, Phys. Rev. D 103, 116009 (2021)

33. D. Acosta, E. Barberis, N. Hurley et al., JINST 18 P09025 (2023)

34. T. von Witzleben et al., Beam Dynamics for Concurrent Operation of the LH*e*C and the HL-LHC, in *Proceedings of 14th International Particle Accelerator Conference* (Venice, Italy, 2023), p. 151